\documentclass[%
preprint,
nobibnotes,
longbibliography,
 amsmath,amssymb,
]{revtex4-1}
\usepackage{dcolumn}
\usepackage{bm}
\usepackage[dvips]{graphicx}
\usepackage{mathrsfs}
\usepackage{amsfonts}
\usepackage{amssymb}
\usepackage{color}
\usepackage{subfigure}
\usepackage{amsmath,latexsym}

\def\d{\mathrm{d}}
\def\eps{\varepsilon}
\newcommand{\lbl}[1]{\label{#1}}

\newcommand{\ignore}[1]{}

\begin{document}

\title{Theoretical analysis for dynamic contact angle hysteresis on chemically patterned surfaces}
\author{Xianmin Xu}%
 \email{Corresponding author, xmxu@lsec.cc.ac.cn}
\affiliation{ LSEC,ICMSEC,
  NCMIS, Academy of Mathematics and Systems Science, Chinese Academy of Sciences, Beijing 100190, China
}
\author{ Xiaoping Wang}%
 \email{mawang@ust.hk}
\affiliation{%
 Department of Mathematics, Hong Kong University of Science and Technology,
Clear Water Bay, Kowloon, Hong Kong, China
}

\begin{abstract}
A dynamic wetting problem is studied for a moving thin fiber inserted in fluid and with a chemically inhomogeneous surface.
A reduced model is derived for contact angle hysteresis by using the Onsager principle
 as an approximation tool. The model is simple and captures the essential dynamics of 
 the contact angle.
From this model we derive an upper bound of the advancing contact angle and
a lower bound of the receding angle, which are verified by numerical simulations. The results are 
 consistent with the quasi-static results. 
The model can also  be used to understand 
 the asymmetric dependence of the advancing and receding contact angles on
the fiber velocity,  which is  observed recently in physical experiments reported in {\it Guan  et al Phys. Rev. Lett.   2016}.
\end{abstract}
\maketitle
\section{Introduction}
Wetting is a common phenomenon in nature and our daily life. It is a fundamental problem
with  applications in many industrial processes, like coating, printing and oil industry, etc.
In equilibrium state, wetting on smooth  homogeneous surfaces can be described  by
the Young's equation \cite{Young1805}. It becomes much more 
 complicated when the solid surface is geometrically rough or chemically  homogeneous \cite{Gennes85,Gennes03,Bonn09,delmas2011contact,giacomello2016wetting}. 
The apparent contact angle on rough surface is not unique even in equilibrium state. 
The largest  contact angle is  called the advancing angle and
the smallest  is called the receding angle. 
The difference between them is  called contact angle hysteresis (CAH).

Theoretical study on the CAH is  difficult due to its multiscale nature.  
The macroscopic contact angles is affected by the microscopic roughness  of the surface.  Previous
studies mainly focus on quasi-static wetting problems on surface with simple geometry or chemical  inhomogeneity (see \cite{Johnson1964,joanny1984model,Schwarts1985,Extrand02,Whyman08,Turco2009,XuWang2011,Hatipogullari19} among many others).
For example, Joanny and De Gennes analysed the CAH on a surface with 
dilute defects \cite{joanny1984model}.  Recently, Hatipogullari {\it et al}  studied the CAH in
a two dimensional chemically heterogeneous microchannel  \cite{Hatipogullari19}.
Similar problems have also been analysed in \cite{XuWang2011}.
For three dimensional wetting problems on surfaces with periodic roughness, there are  only
a few  results. CAH is partly interpreted using the modified 
Wenzel and Cassie equations which are derived to describe the meta-stable states on  rough surfaces \cite{Choi09,Raj12,XuWang2013,xu2016modified}.

The  understanding of CAH in  the dynamic wetting problems is still very poor, although there are some interesting
observations in  experiments\cite{Priest2013,guan2016asymmetric} and molecular dynamics simulations \cite{collet1997dynamics}.
One reason is due to the poor understanding of the  moving contact line problem. Both  modelling
and simulations of  moving contact line (MCL) problems are  challenging in continuum fluid mechanics due to the inherently multiscale nature of MCL \cite{huh1971hydrodynamic,cox1986,Jacqmin2000,QianWangSheng2003,ren2007boundary,
Yue2010,snoeijer2013moving,sui2014numerical,xu2018sharp}. 
When the solid surface is rough
or chemically inhomogeneous, there are only a few numerical studies (e.g. \cite{QianWangSheng2008,RenE2011}) and  theoretical analysis \cite{golestanian2004moving} on the dynamic
wetting problems. 

Recently, a powerful approximation tool is developed by using the Onsager variational principle\cite{onsager1931a,onsager1931b}. 
The main idea is to use the  Onsager  principle 
to derive a reduced model  for 
a set of slow variables. The reduced model is much simpler than the original PDE model and 
it  captures the essential dynamics of
the slow variables. The method has been applied successfully in many
 complicated problems in soft matter and also in hydrodynamics\cite{Doi15,di2018thin,xu2016variational,
DiXuDoi2016,guo2019onset,ManDoi2016,jiang2019application,yu2018capillary,zhou2018dynamics,doi2019application}. 
The Onsager principle is capable to describe the moving contact line problem when the capillary number is small 
and the inertial effect can be ignored. Actually the generalized Navier slip boundary condition for
 contact line motion was derived by using the Onsager principle~\cite{QianWangSheng2006}.

In the paper, we use the Onsager principle as an approximation
tool to study the dynamic wetting problem on chemically inhomogeneous surface. 
We consider the contact line motion on the surface of a thin fiber  inserted in fluid.  
We derive a reduced model  consists of two ODEs for the dynamics of the contact angle and position of  the 
contact line by using the Onsager principle.  The model is easy to analyse
and also easy to solve numerically.  Using  the model, we derive an upper bound for the 
advancing  angle and a lower bound of the receding  angle.
Numerical results show that the model can capture the essential features of
 the dynamic CAH. 
 
We show that the reduced model  characterizes nicely the
asymmetric dependence of the advancing and receding contact angles on
the fiber velocity. The interesting phenomenon has been observed in the recent experiments \cite{guan2016asymmetric}. 
It has been partly analysed by using a phase-field model
in \cite{wang2017dynamic,XuZhaoWang2019}, which is  a quasi-static
model in which  the viscous dissipation
of the fluid is ignored. 
The previous analysis shows that the asymmetric distribution of the 
chemical inhomogeneity on the solid surface may induce the asymmetric dependence
 of the advancing and receding contact angles on the  velocity. Our analysis in this paper 
 reveals that the asymmetric dependence can also be caused by dynamic effects, since 
 the receding contact angle is more sensitive  to the fiber velocity than the advancing angle.
 
The main structure of the paper is as follows. In section 2, we describe the 
main idea of using Onsager principle as an approximation tool
in free boundary problems. In section 3, we use the method to 
derive a  reduced model  for the dynamic wetting problem. In section 4, we show some numerical examples that
demonstrate that the model captures the behavior of the contact line motion and 
 CAH  nicely.
In section 5, we give a few conclusion
remarks.

\section{The Onsager principle as an approximation tool}
The Onsager principle is a variational principle proposed by
Lars Onsager in his celebrated papers on the reciprocal relation\cite{onsager1931a,onsager1931b}.
It has been widely used to derive the time evolving equations in soft matter physics\cite{DoiJPhys,DoiSoftMatter},
such as  the Ericksen-Leslie equation in liquid crystals\cite{de1993physics}
and the gel dynamics equations\cite{doi2009gel} among many others.
In fluid dynamics, it has also been used to derive a generalized Navier Slip boundary 
condition for moving contact line problems\cite{QianWangSheng2006}. 
%


Recently, the Onsager principle has also been used 
as a powerful  tool  to solve approximately many problems in fluid and soft matter systems \cite{Doi15,ManDoi2016,DiXuDoi2016,zhou2018dynamics}.
In particular, some free boundary problems in Stoksean hydrodynamics
can be solved efficiently  by the method \cite{xu2016variational,guo2019onset}.  
The key idea  is described as follows. Suppose we are considering a Stokesian hydrodynamic system 
which includes some free interfaces 
(interfaces between fluid and solid or fluid and fluid).  
The interfaces are moving 
driven by certain potential forces (gravity, surface tension, etc).
Suppose that we are  interested only in the time evolution of 
the free interfaces. 
Let $a(t) = \{a_1(t), a_2(t),..., a_N(t) \} $ be the set of the parameters 
which specify the position of the boundaries.
By ignoring the inertial effect, 
the evolution of the system is determined approximately by using the Onsager principle
for the  parameter set $a(t)$.
The time derivative
$\dot a(t)  = \{ \dot a_1(t),  \dot a_2(t),...  \dot a_N(t) \} $ 
is determined by minimizing the total Rayleighian, which is a function 
of  $ \dot a $ 
\begin{equation}
  R(\dot a, a ) = \Phi (\dot a, a ) + \sum_i \frac{\partial A}{\partial a_i}\dot a_i,
                               \label{eqn:1}
\end{equation}
where $A(a)$ is the potential energy of the system, and  $\Phi (\dot a, a )$
is the energy dissipation function which is defined as the half of 
the minimum of the energy dissipated per unit time in the fluid when the boundary is 
changing at rate $\dot a$.  Since the fluid obeys Stokesian dynamics,  
$\Phi (\dot a, a )$ is always written as a quadratic function of $\dot a$.
\textcolor{black}{
\begin{equation}
  \Phi(\dot a, a ) = \frac{1}{2}\sum_{i,j} \zeta_{ij}(a) \dot a_i \dot a_j.
                               \label{eqn:1a}
\end{equation}
}
The minimum condition of eq.(\ref{eqn:1}) 
\begin{equation}
   \frac{\partial \Phi }{\partial \dot a_i}  +  \frac{\partial A}{\partial a_i} =0
   \quad \mbox{or} \quad 
   \sum_{j} \zeta_{ij}(a) \dot a_j =-  \frac{\partial A}{\partial a_i} ,
                               \label{eqn:2}
\end{equation}
represents the force balance of two kinds of forces, the hydrodynamic frictional force 
$\partial \Phi /\partial \dot a_i$, and the potential force $- \partial A /\partial a_i$ in 
the generalized coordinate. The equation \eqref{eqn:2} gives the dynamics 
of the parameters $a(t)$.

The main  feature of the above approach is that the parameter set $a=\{a_i\}$
may not be a full space to describe the system. It includes only a few slow variables we are interested in.
If $a(t)$ are parameters depending only on time, we are led to a system of ordinary differential equations,
which is much easier to solve and analyse than the standard hydrodynamic equations.
In the following, we will use the idea to study the dynamic CAH
problem on chemically inhomogeneous surfaces.

\section{The analysis of a dynamic wetting problem}
Motivated by the recent physical experiments in \cite{guan2016asymmetric},
we consider a dynamic wetting problem as shown in  Figure~\ref{fig:fiber}. 
A  thin fiber with chemically inhomogeneous surface
is inserted in a liquid reservoir. The liquid-air interface forms a circular contact
line on the fiber surface. By moving the fiber up and down through
the interface with a constant velocity $v$, the contact line moves along the  
surface. When the fiber moves down, the contact line will advance to the upper dry part 
of the fiber surface. This corresponds to an advancing contact angle. Otherwise, if the fiber moves up,
 this corresponds to a  receding contact angle. 
We are interested in the dynamic contact angle hysteresis in the process.
In particular, how the advancing and receding contact angles change with different
velocity $v$. 
\begin{figure}[ht!]
 \centering
  { 
    \includegraphics[width=3.2in]{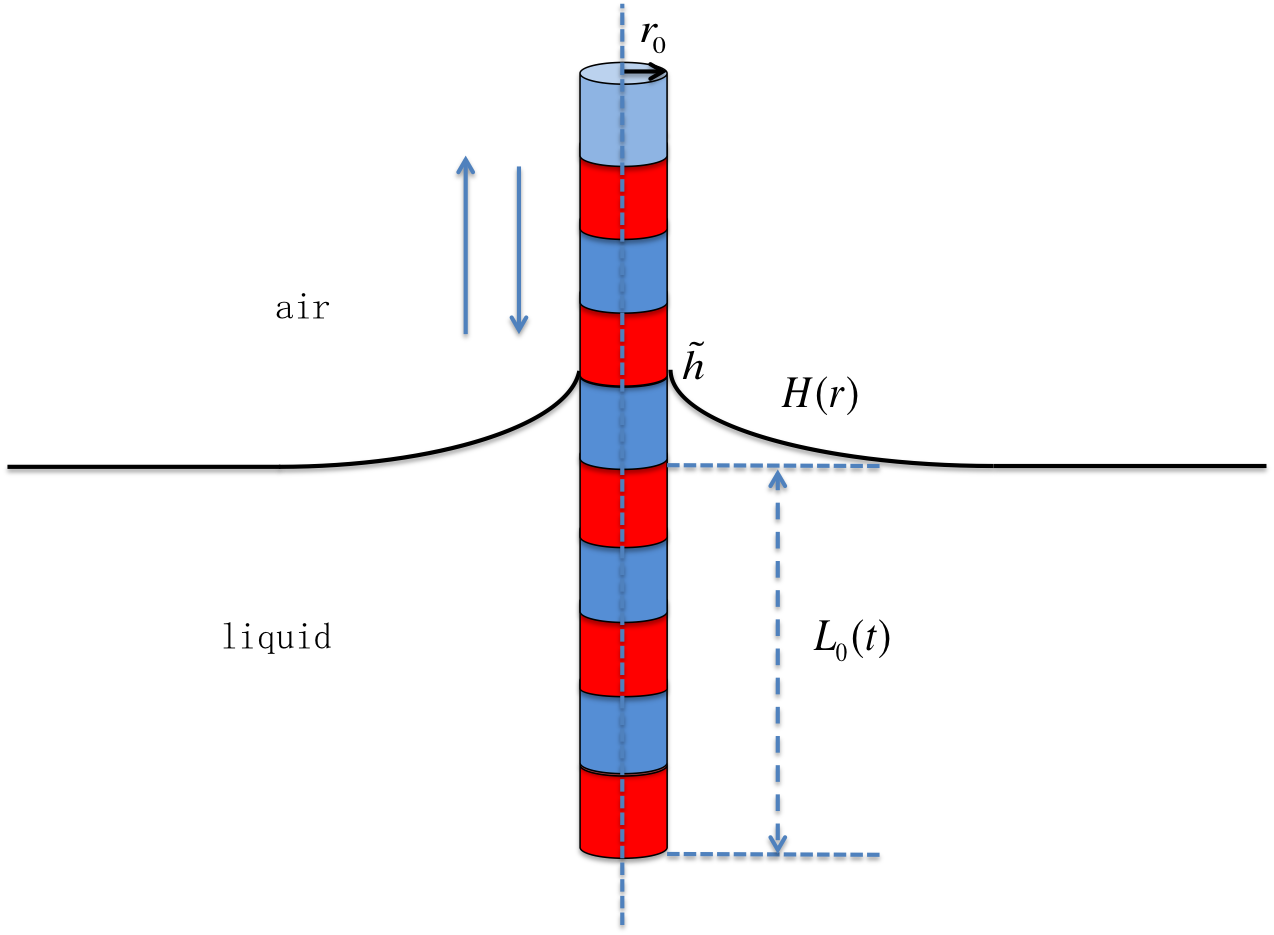}}
  \caption{Fiber with chemically patterned surface in a liquid}
  \lbl{fig:fiber}
\end{figure}

To approximate  the dynamic wetting problem
described above by the Onsager principle, we  make
some ansatz for the system. At time $t$, we assume the liquid-air interface
 is radial-symmetric and described approximately by the  function:
\begin{equation}\label{e:profile}
z=H(r):= h(t)-r_0\cos\theta(t)\ln\Big(\frac{r+\sqrt{r^2-r_0^2\cos^2\theta(t)}}{r_0\cos\theta(t)}\Big), \qquad r\geq r_0.
\end{equation}
Here $r_0$ is the radius of the fiber and it is much smaller than the capillary length $r_c$; the height
$h(t)$ and the dynamic contact angle $\theta(t)$  are two parameters to characterize the evolution of the interface. 
The function~\eqref{e:profile} is a solution of the Young-Laplace equation for the liquid-air meniscus near
a thin cylinder\cite{Gennes03}. The assumption of the profile actually implies that
the liquid-air interface is in local equilibrium away from the contact line. 
This is a good approximation when the velocity of the fiber is relatively small,
i.e. the capillary number is small in the system \cite{guo2019onset,fuentes2019contact}. 
 
Since the gravity will prevent the meniscus from extending indefinitely, we can assume 
the lateral dimension of the meniscus does not exceed the capillary length $r_c$. Suppose the flat part of 
the liquid-air interface is given by $z=0$, then we have
 $H(r_c)=0$. This leads to
\begin{equation}\label{e:relationhtheta}
h(t)=r_0\cos\theta(t)\ln\Big(\frac{r_c+\sqrt{r_c^2-r_0^2\cos^2\theta(t)}}{r_0\cos\theta(t)}\Big).
\end{equation}
It gives a restriction condition between $h(t)$ and $\theta(t)$. In other words, 
the two parameters are not independent. 
We can choose  $\theta(t)$ as the  only slow parameter to characterize 
the evolution of the system. The evolving equation of $\theta(t)$  
will be derived by using the Onsager principle.

\subsection{The surface energy}
We first  calculate the potential energy in the system. Since the length of the meniscus is smaller than
the capillary length, the gravitational energy can be ignored. 
The total potential energy $A$ is composed of some surface energies:
\begin{equation}\label{e:surfaceEnergy}
A=A_{liquid}+A_{fiber},
\end{equation}
where $A_{liquid}$ and $A_{fiber}$ are  the  energy of the liquid-air interface and that
of the fiber surface, respectively.

 Let $L$ be the total length of the fiber and $L_0(t)$ be the length
of the fiber under the horizontal surface $z=0$ at time $t$.
Notice that the position of the contact line is given by
\begin{equation}\label{e:tildeh}
\tilde{h}=h-r_0\cos\theta\ln\big(\frac{1+\sin\theta}{\cos\theta}\big)\approx r_0\cos\theta\ln\big(\frac{2r_c}{r_0(1+\sin\theta)}\big).
\end{equation}
In the approximation, we have used \eqref{e:relationhtheta} and the assumption $r_0\ll r_c$.
 Then we have
\begin{equation}\label{e:SEngfiber}
A_{fiber}=-2\pi\gamma r_0\int_{-L_0(t)}^{\tilde{h}}\cos\theta_Y(z,t)dz.
\end{equation}
 The Young's angle $\theta_Y(z,t)$ 
depends on $z$ and $t$ since the fiber surface is chemically inhomogeneous
and it is moving relative to the horizontal surface $z=0$.
 In addition, since the fiber moves with a velocity $v$, 
  we can assume $\theta_Y(z)=\tilde{\theta}_Y(z-vt)$ for a given function $\tilde{\theta}_Y$, which describes the 
  distribution of the  chemical inhomogeneity on the fiber.

The surface energy $A_{liquid}$ is given by
\begin{equation}\label{e:sEngliquid}
A_{liquid}=2\pi\gamma \int_{r_0}^{r_c}\sqrt{1+(\partial_r H)^2}rdr.
\end{equation}
Direct calculations give 
\begin{equation}\label{e:sEngliquid1}
A_{liquid}=\pi\gamma\Big[r_c\sqrt{r_c^2-r_0^2\cos^2\theta}-r_0^2\sin\theta
+r_0^2\cos^2\theta\ln\Big(\frac{r_c+\sqrt{r_c^2-r_0^2\cos^2\theta}}{r_0(1+\sin\theta)}\Big)\Big].
\end{equation}

To use the Onsager principle, we need compute
the derivative of the total energy with respect to $\theta$. From \eqref{e:SEngfiber}, it is easy to compute
\begin{equation}\label{e:den}
\frac{d A_{fiber}}{d\theta}=\frac{d A_{fiber}}{d \tilde{h}}\frac{d\tilde{h}}{d\theta} 
=-2\pi\gamma r_0^2\cos\tilde{\theta}_{Y}(\tilde{h}-vt) g(\theta),
\end{equation}
with
\begin{equation}\label{e:fun_g}
g(\theta)=r_0^{-1}\frac{d \tilde{h}}{d \theta}\approx -\big[\sin\theta\ln\big(\frac{2 r_c}{r_0(1+\cos\theta)}\big)+1-\sin\theta\big].
\end{equation}
By direct calculations from \eqref{e:sEngliquid1}, we obtain
\begin{align*}
\frac{d A_{liquid}}{d\theta}=&\pi \gamma r_0^2\Big[\frac{r_c\cos\theta\sin\theta}{\sqrt{r_c^2-r_0^2\cos^2 \theta}}-\cos\theta
-2\sin\theta\cos\theta\ln\Big(\frac{r_c+\sqrt{r_c^2-r_0^2\cos^2\theta}}{r_0(1+\sin\theta)}\Big) \\
 &+\frac{\cos^2\theta}{r_c+\sqrt{r_c^2-r_0^2\cos^2\theta}}\cdot\frac{r_0^2\sin\theta\cos\theta}{\sqrt{r_c^2-r_0^2\cos^2\theta}}
 -\frac{\cos^3\theta}{1+\sin\theta} \Big]\\
 \approx &\pi\gamma r_0^2\cos\theta\Big[\sin\theta-1-2\sin\theta\ln\big(\frac{2 r_c}{r_0(1+\sin\theta)}\big)+\frac{r_0^2}{2r_c^2}\cos^2\theta\sin\theta-\frac{\cos^2\theta}{1+\sin\theta}\Big]\\
 \approx &2\pi \gamma r_0^2\cos\theta g(\theta).
\end{align*}
Here we use $r_0\ll r_c$ in the above approximations. 
Combining the above analysis, we obtain 
\begin{equation}\label{e:dSurfEng}
\frac{d A}{d \theta}=\frac{d A_{liquid}}{d\theta}+\frac{d A_{fiber}}{d\theta}\approx 2\pi\gamma r_0^2 g(\theta)\big[\cos\theta-\cos\tilde{\theta}_Y(\tilde{h}-vt)\big].
\end{equation}
This is a general force which makes the dynamic contact angle $\theta$
to relax to its equilibrium value (the Young's angle).

\subsection{The energy dissipation}
We then compute the viscous energy dissipations in the system. Since the capillary number is small, we can
assume the viscous dissipations near the contact line is dominant in the system\cite{Gennes03}. 
When the contact angle $\theta$ is small,
the Rayleigh dissipation function is approximately given by \cite{Gennes03}
\begin{equation}\label{e:diss_smallAng}
\Phi=\frac{3\pi\eta r_0}{\theta}|\ln\eps|U^2,
\end{equation}
where $U$ is the slip velocity of the contact line on the fiber and $\eps$ is a cut-off parameter to avoid the singular integration.
The formula \eqref{e:diss_smallAng} can be  derived by computing the
viscous energy dissipation in a two dimensional wedge region  as shown in Figure~\ref{fig:wedge} by
 lubrication approximations. It works only when $\theta\ll 1$.

\begin{figure}[ht!]
 \centering
  { 
    \includegraphics[width=2.3in]{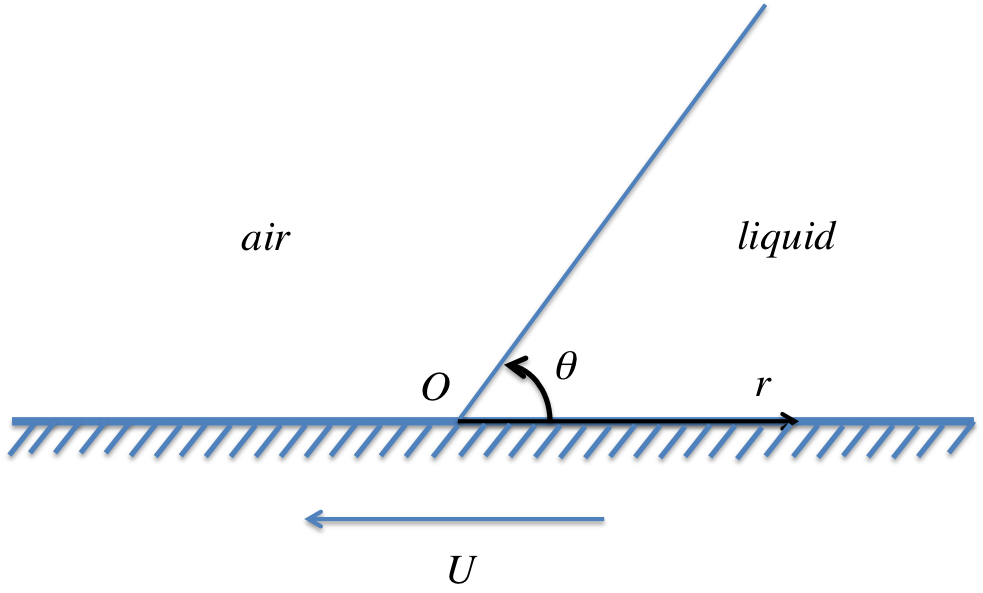}}
  \caption{Fluid near a contact line}
  \lbl{fig:wedge}
\end{figure}
When the contact angle $\theta$ is large, we need  an alternative calculations for
 the viscous dissipations. Here we adopt the method  in \cite{huh1971hydrodynamic}.
We first compute the dissipation in the two-dimensional wedge region.
As shown in Figure~\ref{fig:wedge}, we choose a polar  coordinate system. The
origin $O$ is set at the contact point. The liquid region is given by 
$\{(r,\phi)| r>0,0<\phi<\theta\}.$
Suppose the solid surface moves with a velocity $U$. 
The velocity field of the liquid is described by the Stokes equation.
By the incompressibility condition, we can define a stream function
\begin{equation}\label{e:streamfunc}
\psi(r,\phi)=r(a\sin\phi+b\cos\phi+c\phi\sin\phi+d\phi\cos\phi),
\end{equation}
where $a,b,c$ and $d$ are parameters to be determined.
Then the velocities in the radial and angular directions are given by
\begin{equation}\label{e:velocity}
v_r=-\frac{1}{r}\partial_{\phi}\psi, \quad 
v_{\phi}=\partial_r\psi.
\end{equation}
The boundary conditions of the fluid equation are
\begin{equation}\label{e:bnd}
\left\{
\begin{array}{ll}
\partial_r\psi=0& \hbox{on }\phi=0 \hbox{ and } \phi=\theta, r>0,\\
-\frac{1}{r}\partial_{\phi}\psi=U& \hbox{on } \phi=0, r>0,\\
\partial_{\phi\phi}\psi=0& \hbox{on } \phi=\theta, r>0.
\end{array}
\right.
\end{equation}
Here we adopt a no-slip boundary condition on the solid surface and set the normal velocity and the tangential stress to be zero
on the liquid-air interface. 
By substituting the equation \eqref{e:streamfunc} to the boundary conditions \eqref{e:bnd}, we  obtain
\begin{equation}\label{e:notations}
a=-\frac{\theta U}{\theta-\sin\theta\cos\theta},\quad
b=0,\quad
c=\frac{\sin^2\theta U}{\theta-\sin\theta\cos\theta},\quad
d=\frac{\sin\theta\cos\theta U}{\theta-\sin\theta\cos\theta}
\end{equation}
Then the equation \eqref{e:velocity} leads to
\begin{equation}\label{e:velocity1}
\left\{\begin{array}{l}
v_r=-\big[(a+d)\cos\phi+c\sin\phi+c\phi\cos\phi-d\phi\sin\phi\big],\\
v_{\phi}=a\sin\phi+c\phi\sin\phi+d\phi\cos\phi,
\end{array}
\right.
\end{equation}
where $a,c,d$ are given in \eqref{e:notations}.
Let $\mathbf{r}$ and $\mathbf{n}$ be the unit vectors along the radial and angular
directions, respectively. Direct computations yield
\begin{equation*}
\nabla \mathbf{v}
=\frac{2}{r}(d\sin\phi-c\cos\phi)\mathbf{n}\mathbf{r}^{T}.
\end{equation*}
Then the total viscous energy dissipation in the two dimensional wedge (liquid) region can be computed out as
\begin{align*}
\Psi=\int\int \eta|\nabla \mathbf{v}|^2 r\d\phi d r=\frac{2\eta |\ln\eps|\sin^2\theta U^2}{\theta-\sin\theta\cos\theta},
\end{align*}
where $\eps$ is the cut-off parameter.  

Using the above analysis result, the energy dissipation in our system can be approximated by
$4\pi r_0 \Psi.$
Then the  Rayleigh dissipation function $\Phi$,
which is defined as half of the total energy dissipation, is given by
\begin{equation}\label{e:diss}
\Phi=2\pi\eta r_0|\ln \eps|\frac{\sin^2\theta}{\theta-\sin\theta\cos\theta} U^2.
\end{equation}
The formula is consistent with the equation \eqref{e:diss_smallAng} when the contact angle is small. 
Actually, when $\theta\ll 1$, 
$\sin\theta\sim \theta-\frac{\theta^3}{6}+\cdots$,  $\cos\theta=1-\frac{\theta^2}{2}+\cdots$,  
we easily have  $\frac{\sin^2\theta }{\theta-\sin\theta\cos\theta}\sim\frac{3}{2\theta}.$
The equation \eqref{e:diss} will reduce to  \eqref{e:diss_smallAng}.

In Equation \eqref{e:diss}, $U$ is the relative velocity of between the fiber and the  contact line.
 In our case, it is written as 
$$U=\dot{\tilde{h}}-v=\frac{d \tilde{h}}{d \theta}\dot{\theta}-v=r_0 g(\theta)\dot{\theta}-v,$$
where we have used \eqref{e:fun_g}. Then the Rayleigh dissipation function is 
\begin{equation}\label{e:dissfun}
\Phi=\frac{2\pi\eta r_0|\ln \eps|\sin^2\theta}{\theta-\sin\theta\cos\theta} \big(r_0 g(\theta)\dot{\theta}-v\big)^2.
\end{equation}
We can further compute
\begin{equation}\label{e:D_dissfun}
\frac{d \Phi}{d\dot{\theta}}=\frac{4\pi\eta r_0^2|\ln \eps|\sin^2\theta}{\theta-\sin\theta\cos\theta} g(\theta)\big(r_0 g(\theta)\dot{\theta}-v\big).
\end{equation}

\subsection{The evolution equation  for the dynamic contact angle}
By using the Onsager principle  for the dynamic contact angle, we have
$\frac{\partial\Phi}{\partial\dot{\theta}}+\frac{\partial A}{\partial\theta}=0$.
Combining with the previous analytic results, we obtain a dynamic equation for the contact angle $\theta$,
\begin{equation}\label{e:temp}
\frac{2\eta|\ln\eps|\sin^2\theta}{\theta-\sin\theta\cos\theta}(r_0g(\theta)\dot{\theta}-v)+\gamma \big[\cos\theta-\cos\tilde{\theta}_Y(\tilde{h}-vt)\big]=0.
\end{equation}
Introduce a notation
\begin{equation}\label{e:fun_f}
f(\theta)=\frac{\theta-\sin\theta\cos\theta}{2|\ln\eps|\sin^2\theta}.
\end{equation}
The equation \eqref{e:temp} is simplified to
\begin{equation}\label{e:dynEqAng}
\dot{\theta}={(r_0 g(\theta))^{-1}}\big[v-v^*f(\theta)(\cos\theta-\cos\tilde{\theta}_Y(\tilde{h}-vt))\big].
\end{equation}
where $v^*=\frac{\gamma}{\eta}$ and $\tilde{h}$ is given in \eqref{e:tildeh}.  This is an 
ordinary differential equation for $\theta$ and can be solved easily by standard numerical methods. 

For convenience in applications, we can rewrite \eqref{e:dynEqAng} into a different equivalent form.
Introduce a new variable $Z=\tilde{h}-vt$, which
represents the position of the contact line relative to the fiber surface. 
By \eqref{e:fun_g} and \eqref{e:dynEqAng}, we can compute 
\begin{equation*}
\frac{d Z}{d t}=\frac{d \tilde{h}}{d \theta}\dot{\theta} -v =r_0 g(\theta)\dot{\theta}-v=-v^* f(\theta)\big[\cos\theta-\cos\tilde{\theta}_Y(Z)\big].
\end{equation*}
The equation \eqref{e:dynEqAng} is equivalent to an ordinary differential system for $\theta$ and $Z$:
\begin{equation}\label{e:ODE}
\left\{
\begin{array}{l}
\dot{\theta}={(r_0 g(\theta))^{-1}}\left[v^* {f}(\theta)(\cos\tilde{\theta}_Y(Z)-\cos\theta)+v \right],
\\
\dot{Z}=v^*{f}(\theta)(\cos\tilde{\theta}_Y(Z)-\cos\theta),
\end{array}
\right.
\end{equation}
where the dimensionless notations $f(\theta)$ and $g(\theta)$ are given in \eqref{e:fun_f} and \eqref{e:fun_g}, respectively.
The structure of the system \eqref{e:ODE} is similar to a model derived from the analysis of the 
phase-field equation in \cite{wang2017dynamic}, where
the viscous dissipation is  ignored.

\subsection{Discussions}In general the ODE system \eqref{e:ODE} can not be solved explicitly.
But we can do some simple analysis to give some physical understandings for the equation.
 We first consider the case when the fiber surface is homogeneous.
 We can assume $\tilde{\theta}_Y\equiv \theta_0$ for a constant $\theta_0$. 
 Then the two equation in \eqref{e:ODE} are decoupled. 
 The evolution of the dynamic contact angle $\theta$ can be  determined solely by the first equation.
It is easy to see that there exists a steady state, in which the contact angle does not change with time,  when 
\begin{equation}\label{e:steadystate}
v=v^* {f}(\theta)(\cos\theta -\cos {\theta}_0).
\end{equation}
The equation gives a relation between the capillary number $Ca=v/v^*$ and the  dynamic contact angle $\theta$.
It can be rewritten as
\begin{equation}\label{e:dynamicAngle}
\theta=G(\theta_0,Ca),
\end{equation}
where $G$ is an implicit function.
This implies that the dynamic contact angle in steady state is determined the Young's angle and the capillary number.
By solving the nonlinear algebraic equation \eqref{e:steadystate} numerically, 
 we draw a curve for $\theta=G(\theta_0,Ca)$ for $\theta_0$ in Figure~\ref{fig:steady}.
The curve shows that the dynamic contact angle in steady state is a monotone function with respect to $Ca$. 
The equilibrium contact angle is equal to the Young's angle $\theta_0$ when $Ca=0$. When $Ca>0$(the fiber moves up),
 the dynamic contact angle corresponds to a receding angle such that $\theta<\theta_0$.
 When $Ca<0$(the fiber moves down), $\theta$ corresponds to  an advancing 
contact angle which is larger than $\theta_0$. The results are consistent with those in previous analysis \cite{cox1986,Gennes03,xu2016variational}.
\begin{figure}[ht!]
 \centering
    \includegraphics[width=3.4in]{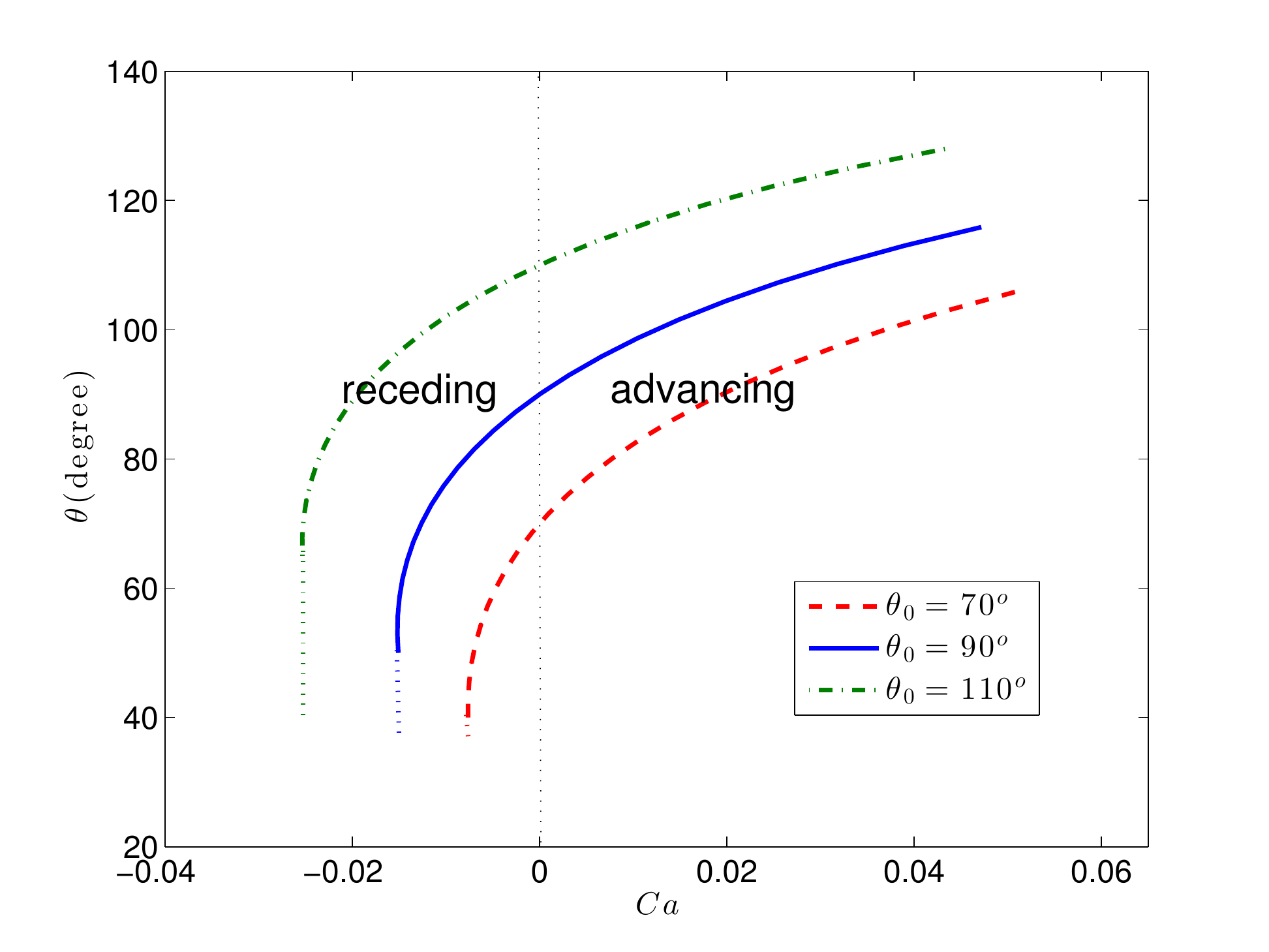}
  \caption{The relation between the dynamic contact angle and the Capillary number in steady state.}
  \lbl{fig:steady}
\end{figure}
%

When the fiber surface is chemically inhomogeneous,  it is  more difficult
to analyze for the equation~\eqref{e:ODE}. We consider a special case where
the surface is composed by two materials with different Young's angles
 $\theta_{Y1}$ and $\theta_{Y2}$ (suppose $\theta_{Y1}>\theta_{Y2}$). 
 The two materials are distributed periodically on the surface.
 If the period is large, the motion of the contact line on one material is like that on
 a homogeneous surface. The dynamic contact angle may alternate between two different steady states.  
The corresponding dynamic contact angles are $\theta_1=G(\theta_{Y1},Ca)$
and $\theta_2=G(\theta_{Y2},Ca)$, respectively. Since $\theta_{Y1}>\theta_{Y2}$,
$\theta_1$ gives an upper bound for the contact angle  while 
$\theta_2$ gives a lower bound. In the next section, we will verify this numerically.
When $Ca$ goes to zero, $\theta_{1}$ and $\theta_2$ will  converge to $\theta_{Y1}$ and
$\theta_{Y2}$, respectively. This is consistent with the previous analysis for
the quasi-static CAH in \cite{XuWang2011,Hatipogullari19}.

\section{Numerical results}
We  show some numerical results for the equation~\eqref{e:ODE} for some chemically inhomogeneous
fiber surfaces. 
 We assume that $\tilde{\theta}_Y(z)$ is a period function with a period $l$ such that $\delta=l/ r_c\ll 1$.
 Then we can write $
 \tilde{\theta}_Y(z)= {\vartheta}_Y(\frac{z}{l}),
 $ 
with a function ${\vartheta}_Y(\cdot)$ being a function with period $1$.

We  non-dimensionalize  the equation~\eqref{e:ODE}. Suppose the characteristic  velocity  is given by
the  velocity $v^*=\frac{\gamma}{\eta}$ and the characteristic length scale is the capillary length $r_c$.
Then the characteristic time is given by $t_c=r_c/v^*$. 
Introduce some dimensionless parameters $\delta_0=\frac{r_0}{r_c}$ and $Ca=\frac{v}{v^*}$,
and  use $\tilde Z$ to represent the dimensionless coordinate $\frac{Z}{r_c}$, then the system 
\eqref{e:ODE} could be rewritten as
\begin{equation}\label{e:ODE1}
\left\{
\begin{array}{l}
\dot{\theta}={(\delta_0  {g}(\theta))^{-1}}\left[ {f}(\theta)(\cos {\vartheta}_Y(\frac{\tilde Z}{\delta})-\cos\theta)+Ca \right],
\\
\dot{\tilde Z}={f}(\theta)(\cos {\vartheta}_Y(\frac{\tilde Z}{\delta})-\cos\theta).
\end{array}
\right.
\end{equation}
Here the function $f(\theta)$ is given in \eqref{e:fun_f} and $g(\theta)$ can be rewritten as
 $${g}(\theta)=-\sin\theta\ln\Big(\frac{2\delta_0^{-1}}{1+\sin\theta}\Big)+\sin\theta-1.$$ 
 For simplicity in notations, we still use $Z$ to represent the dimensionless $\tilde{Z}$ hereinafter.


The equation \eqref{e:ODE1} can be solved easily by the standard Runge-Kutta method. 
We set $\delta_0=4.03\times 10^{-4}$ which is chosen from \cite{guan2016asymmetric}. 
The cut-off parameter is chosen to satisfy $\ln \eps=13.8$, which is a typical value used 
in literature\cite{Gennes03}. 
 In the following, we show some numerical results for some special choices of ${\vartheta}_Y(z)$.

In the first example, we choose $\tilde{\theta}_Y(z)=110^o+8^o\cdot \sin(2\pi z/\delta)$. 
This is a smooth periodic function.
The maximal Young's angle is $\theta_{Ymax}=118^o$ and minimal Young's angle is $\theta_{Ymin}=102^o$.

We first set $|Ca|=0.0001$ and test for different  period $\delta$. 
In this case, the
capillary number is very small so that it is very close to a 	quasi-static process.
Some typical numerical results are shown in the first two subfigures of Figure~\ref{fig:hysteresisdelta}.
They are the trajectories of the solution of \eqref{e:ODE1} in the phase space.
Both the dynamic contact angle and receding contact angles are shown with respect to $Z$.
When the oscillation of the Young's angle $\tilde{\theta}_Y$  is weak ($\delta=0.003$), 
the advancing and receding cases have almost
the same trajectory in the phase space. There seems no hysteresis.
When $\delta$ decreases, the stronger oscillation of $\tilde{\theta}_Y$ corresponds to stronger chemical inhomogeneity.
Then  the advancing and receding processes may follow different trajectories. When $\delta=0.0003$, 
there is obvious
CAH phenomenon.  The advancing and receding contact angles oscillate around
different values. 
The largest dynamic contact angle is about $118.02^o\approx G(\theta_{Ymax},-|Ca|)$
and the smallest contact angel is about $101.9^o\approx G(\theta_{Ymin},|Ca|)$. Here $G$ 
is an implicit function given in \eqref{e:dynamicAngle}. Since the capillary number is very small,
the two angles  are  very close  to the largest Young's and smallest Young's angles in the system.

We then set $|Ca|=0.0025$ and test for different $\delta$. 
Some typical numerical results are shown in the last two subfigures of Figure~\ref{fig:hysteresisdelta}.
We can see that when the inhomogeneity is relatively weak ($\delta=0.003$), the advancing 
and receding processes follow different but partially overlapped trajectories.
The largest advancing contact angle is about $119.33^o\approx G(\theta_{Ymax},-|Ca|)$
and the smallest receding contact angle is about $99.12^o\approx G(\theta_{Ymin},|Ca|)$.
When the chemical inhomogeneity becomes strong enough (e.g. $\delta=0.0003$), 
 the trajectories separate completely for the 
advancing and receding cases. This corresponds to obvious CAH.
The largest advancing contact angle is slightly smaller than the upper bound
$119.33^o$ and the advancing contact angle is  slightly larger than the lower bound $99.12^o$.
This is because  the strong inhomogeneity and the large velocity of the fiber
make the motion of the contact angle away from a steady state.

\begin{figure}[ht!]
 \centering
    \includegraphics[width=2.7in]{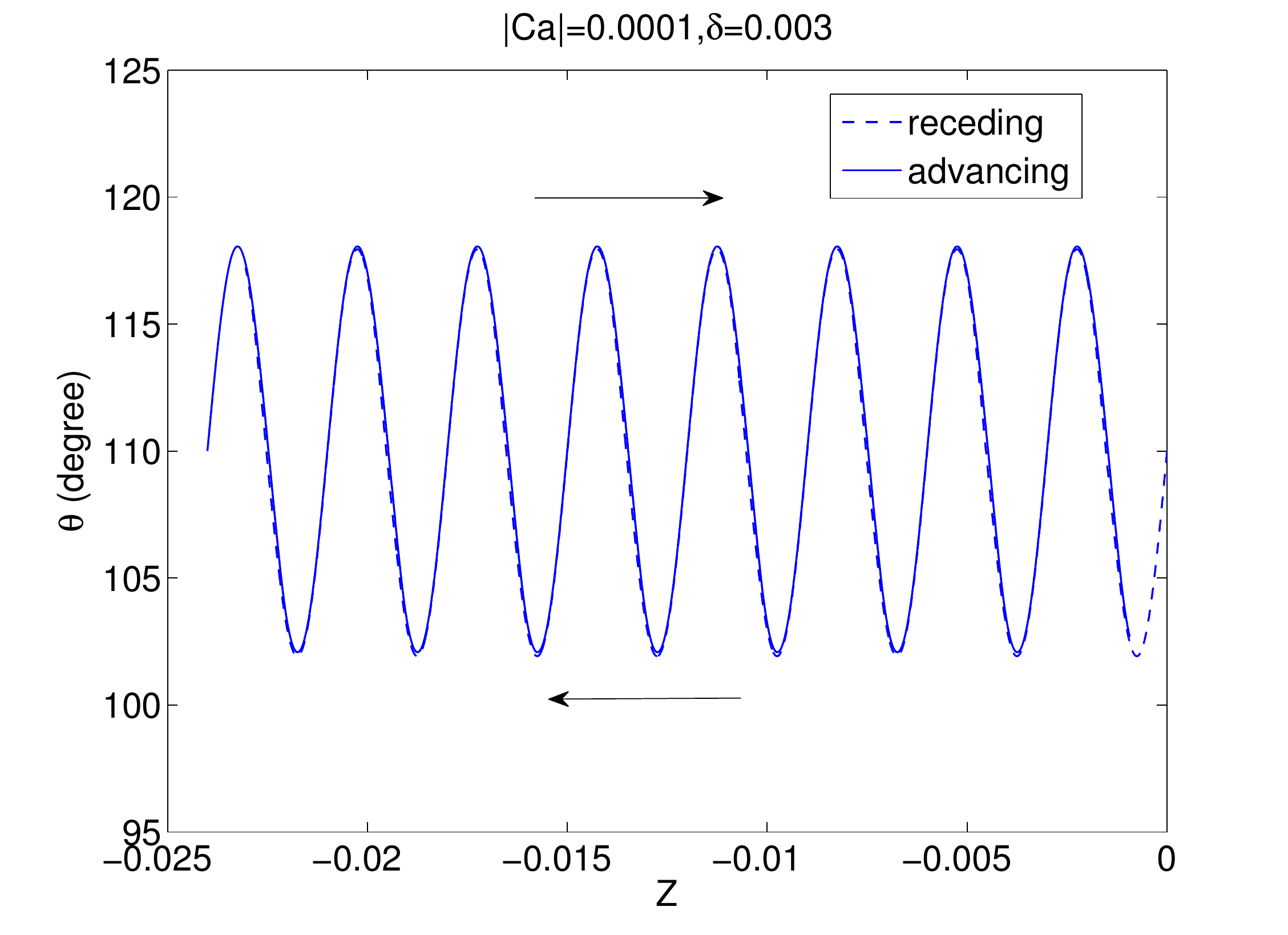}
    \includegraphics[width=2.7in]{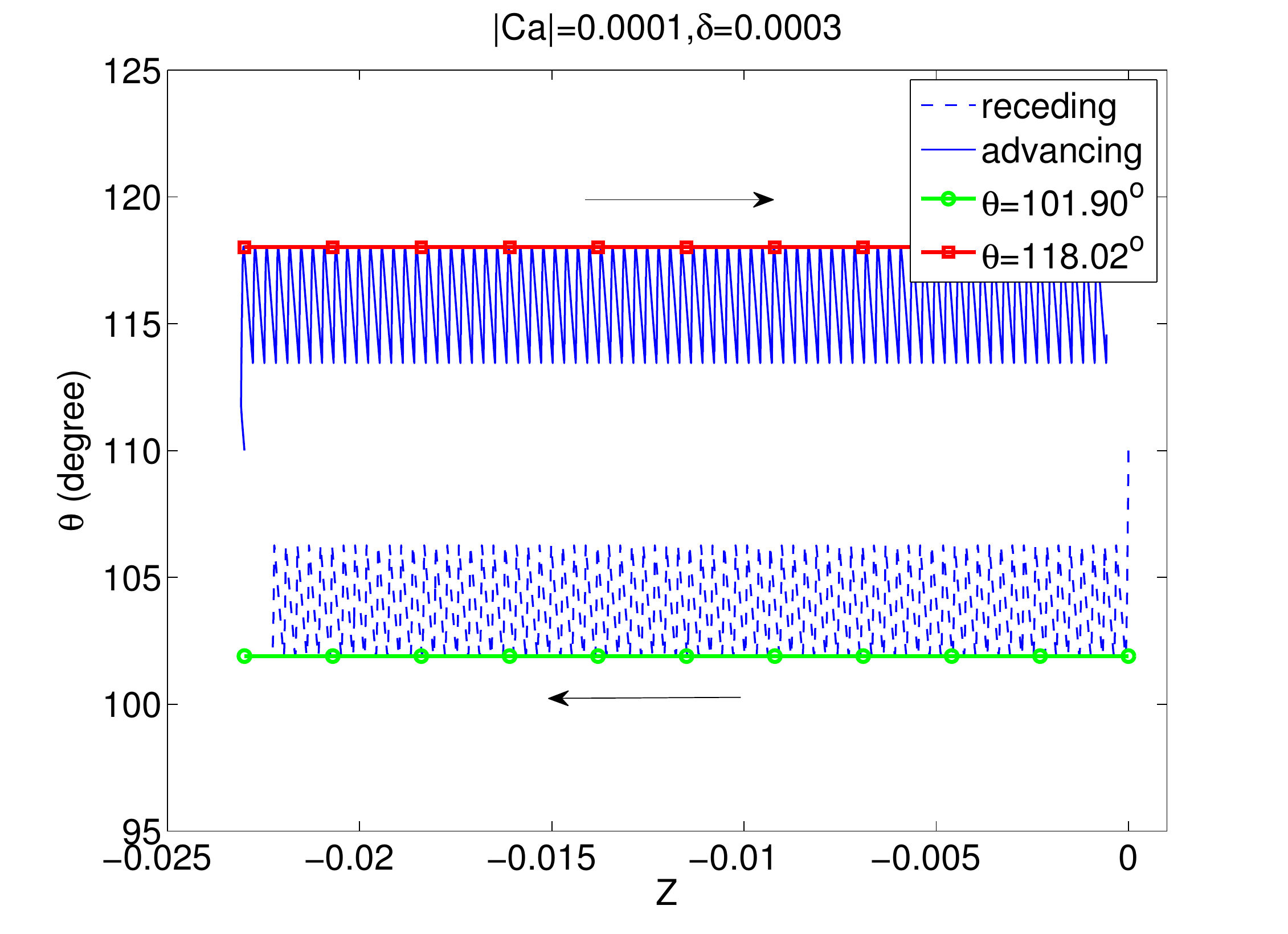}
    \includegraphics[width=2.7in]{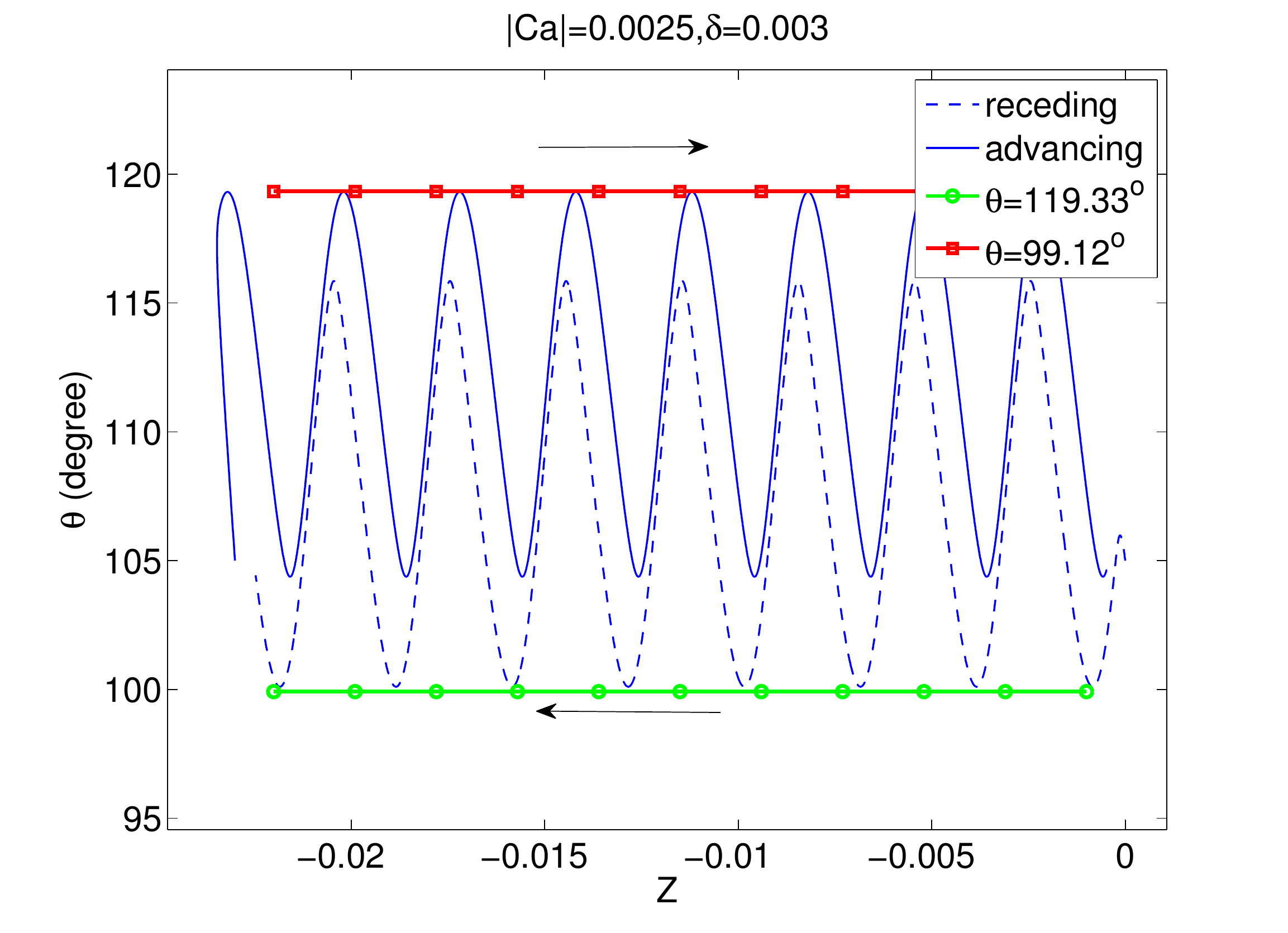}
    \includegraphics[width=2.7in]{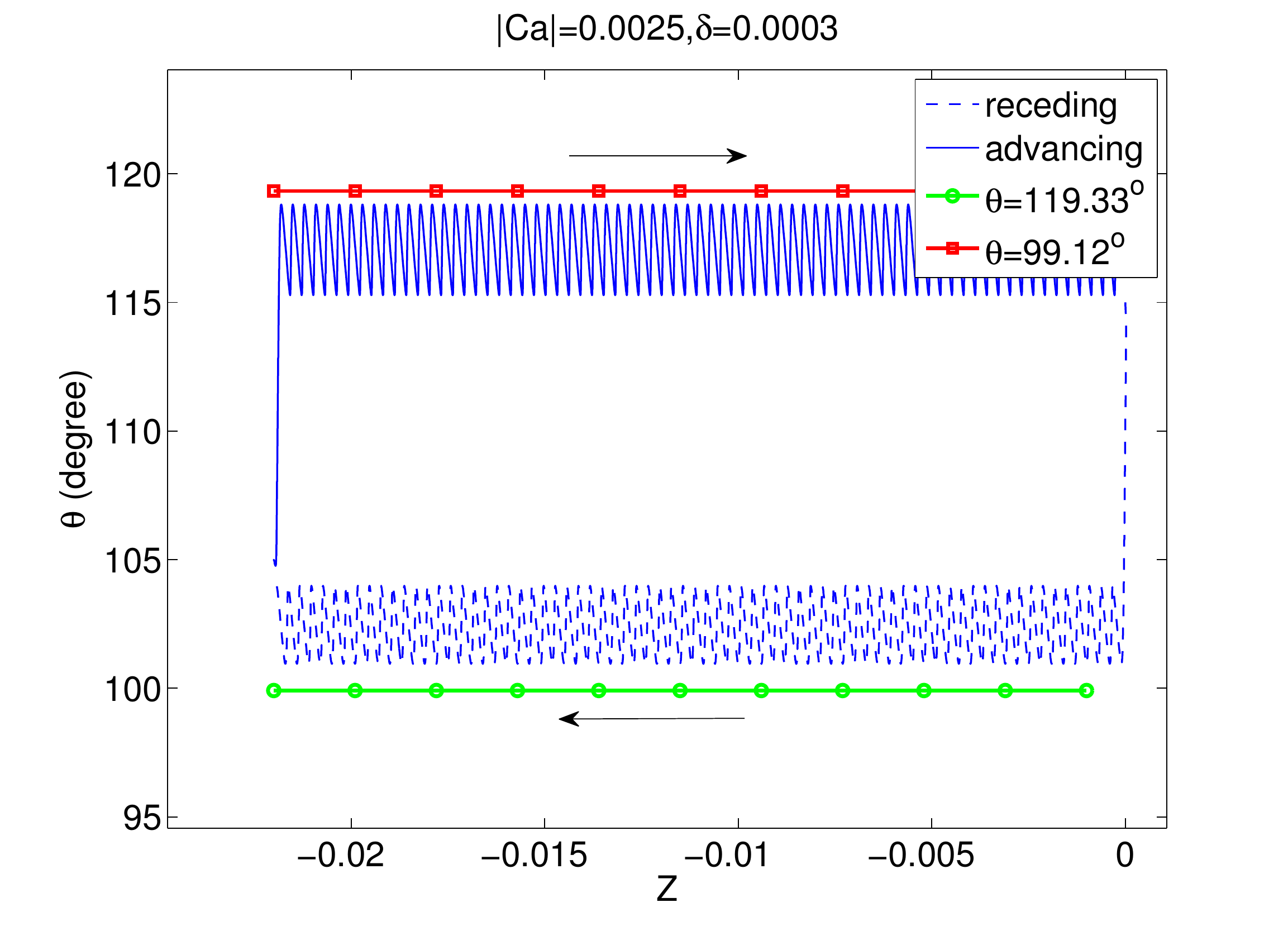}
  \caption{Trajectories of the dynamic contact angles and the contact line positions in phase plane for different $\delta$ and $Ca$. (Example 1)}
  \lbl{fig:hysteresisdelta}
\end{figure}

In the second example, we set $\tilde{\theta}_Y(z)=100^o+20^o\cdot \tanh(10\sin(2\pi z/\delta))$. 
This approximates  a  chemically patterned surface with different Young's angles $\theta_A=120^o$ and 
$\theta_B= 80^o$.  

We first set $|Ca|=0.0001$ and test for different $\delta$. 
Some typical numerical results are shown in the first two subfigures in Figure~\ref{fig:hysteresisZig}.
We  see that  the advancing and receding trajectories do not coincide even for very large period $\delta$
(e.g $\delta=0.03$). This is different from the case when $\theta_{Y}$ is a smooth function as in the first example.
The phenomena are like that the pinning of the contact line by a single defect in quasi-static processes\cite{joanny1984model,delmas2011contact}.
When $\delta$ becomes small enough, the two trajectories  separate completely. 
When $\delta=0.0003$,
the advancing and receding contact angles oscillate around different values. The
largest contact angle is almost equal to the upper bound $120.00^o\approx G(\theta_A,-|Ca|)$ and the smallest contact angle is 
almost equal to the lower one  $79.83^o\approx G(\theta_B,|Ca|)$. Since $|Ca|$ is small, the two values
 are close to the two Young's angles.
The results are consistent to the previous analysis for  the quasi-static CAH on chemically patterned surface\cite{XuWang2011,Hatipogullari19}. 

We then set $|Ca|=0.0025$ and test for different periods  $\delta$.
Some typical numerical results are shown in  the last two subfigures in Figure~\ref{fig:hysteresisZig}. We can observe
  clear stick-slip behaviour for $\delta=0.003$. The stick behaviour corresponds to the case that the position of the contact line 
  does not change much while the contact angle change dramatically. The slip behaviour corresponds to
  a process that both the contact line and the contact angle change dramatically(the slope parts on the trajectories).
    The advancing and receding processes have
   different but slightly overlapped trajectories. The largest advancing angle almost equal to the upper bound $121.29^o=G(\theta_A,-|Ca|)$
  and the smaller receding angle is almost the lower bound $76.52^o=G(\theta_B,|Ca|)$. 
  When the contact angles reaches the upper and lower bounds, there exist 
  some steady states that the contact angles does not change while the contact line moves.
   When $\delta=0.0003$, the advancing and receding trajectories separate completely. 
The advancing contact angle is slightly smaller than
 the upper bound ($121.29^o$) and the receding contact angle is slightly larger than the lower one
 ($76.52^o$).
 This is because there is no steady state in the  advancing or receding processes.
 However, the deviation is smaller than that in the previous example(comparing with the last two subfigures in Figure~\ref{fig:hysteresisdelta}).
\begin{figure}[ht!]
 \centering
    \includegraphics[width=2.7in]{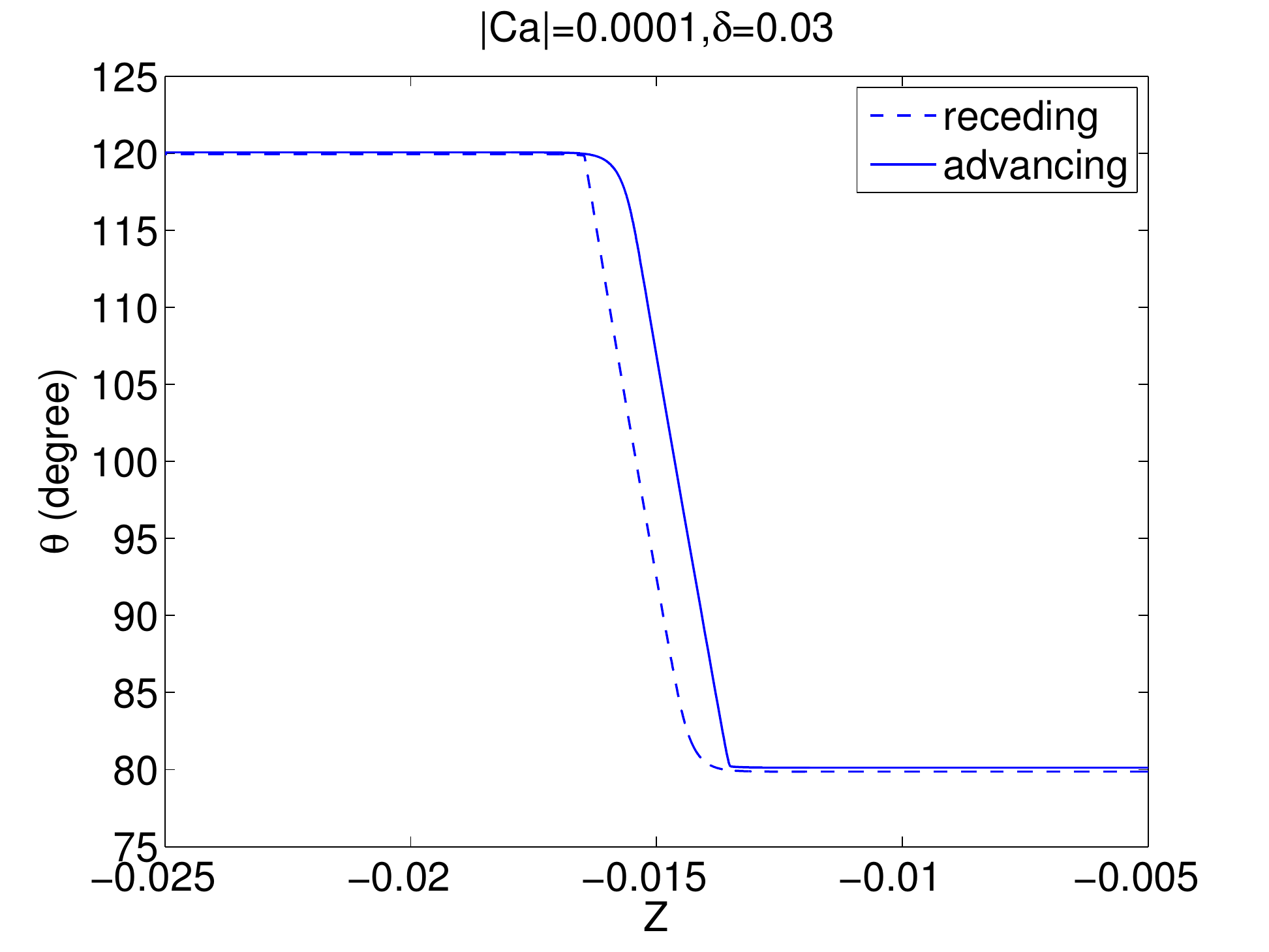}
    \includegraphics[width=2.7in]{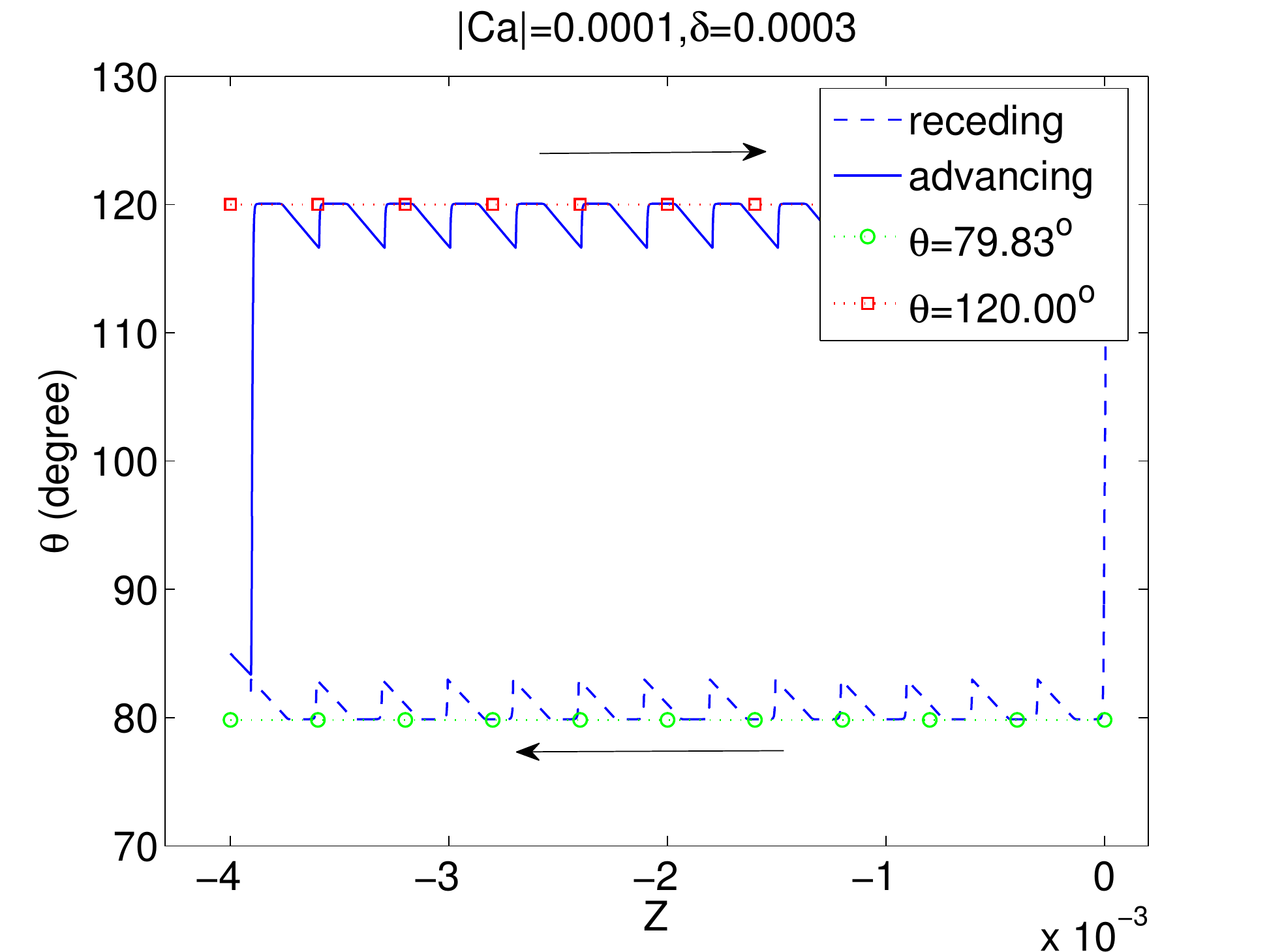}
    \includegraphics[width=2.7in]{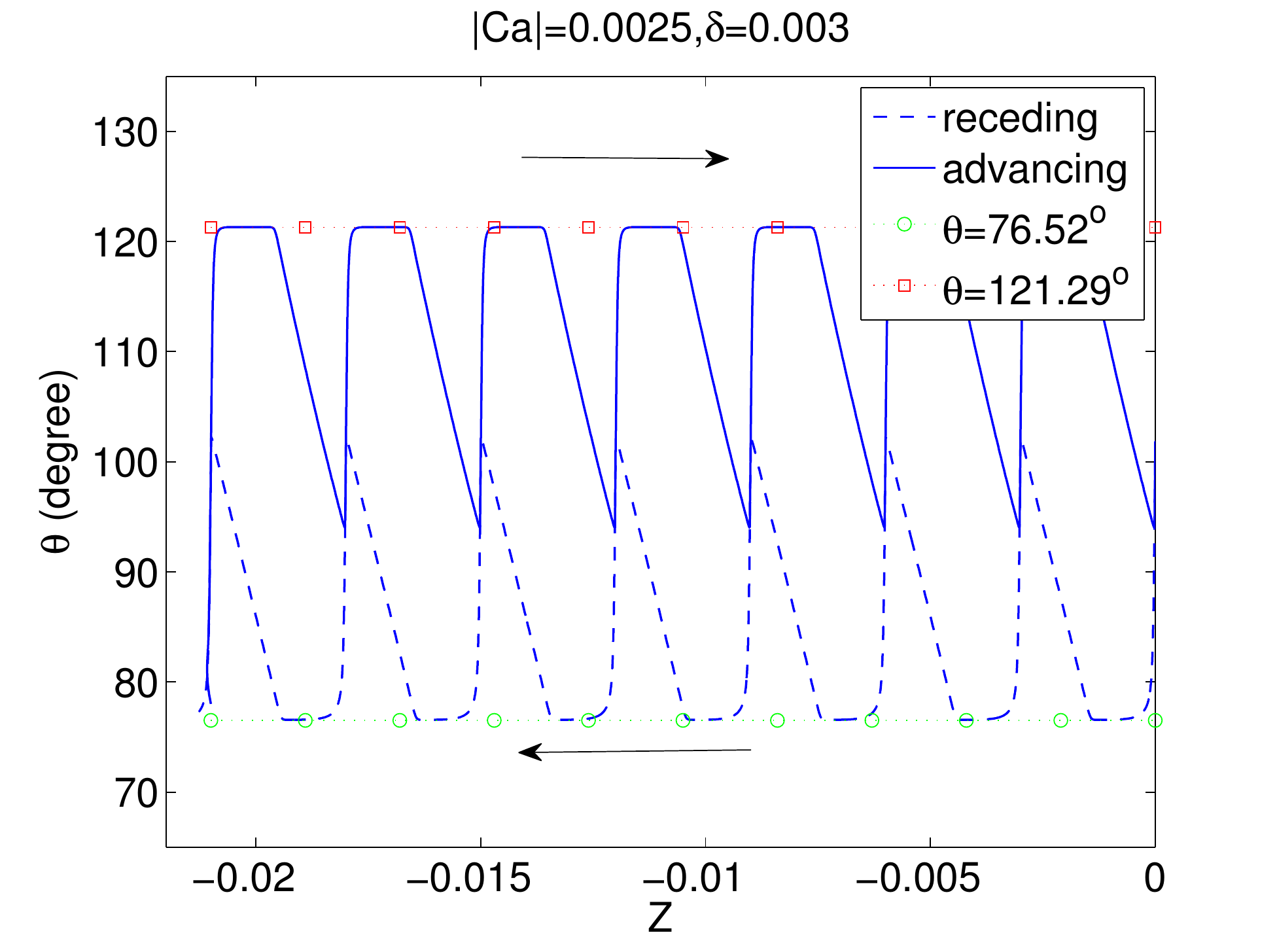}
    \includegraphics[width=2.7in]{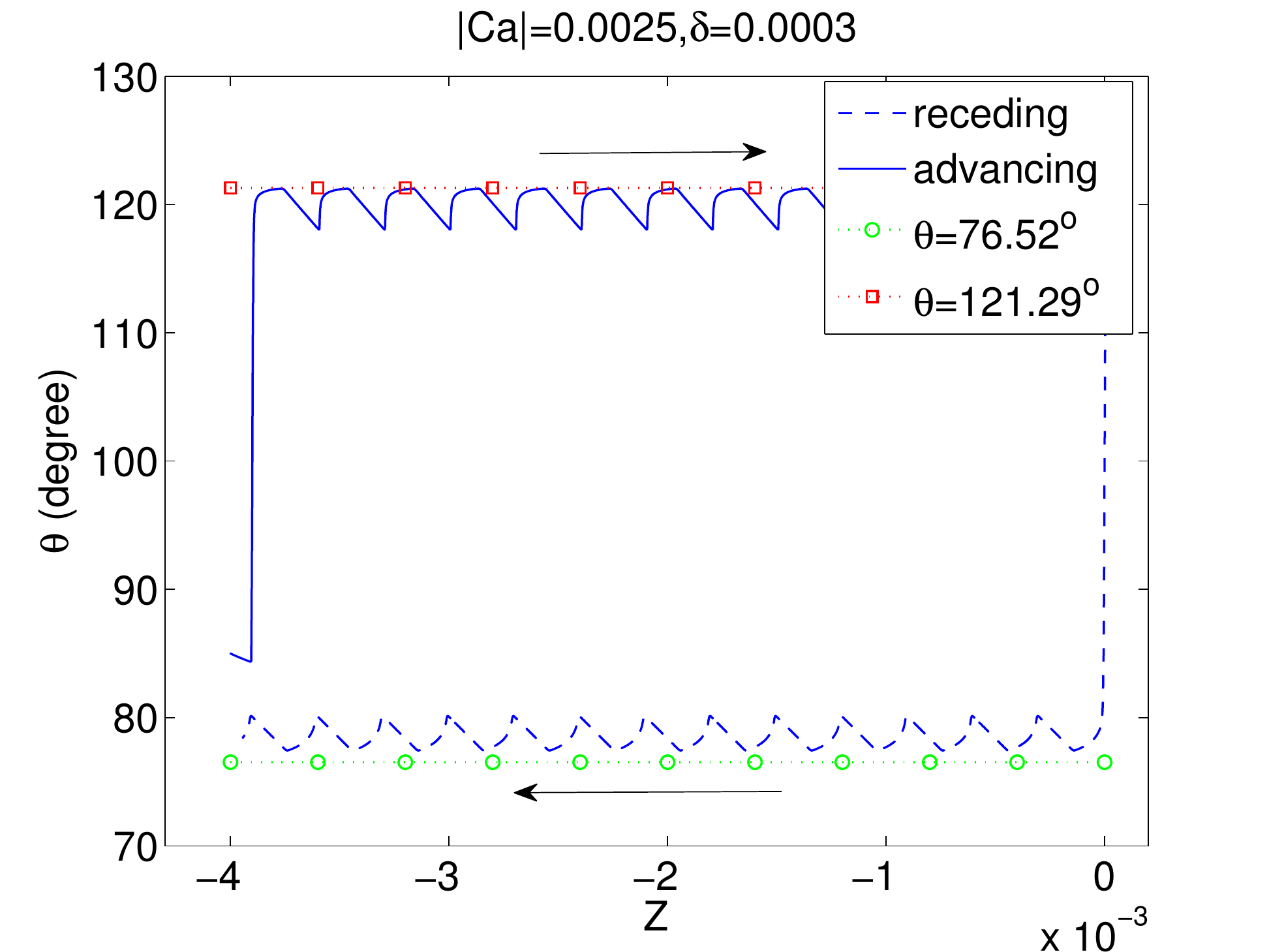}
  \caption{Trajectories of the dynamic contact angles and the contact line positions  in phase plane for different $\delta$ and $Ca$. (Example 2)}  \lbl{fig:hysteresisZig}
\end{figure}

In the last example, we investigate in more details on how the dynamic CAH be affected the fiber velocity.
We choose $\tilde{\theta}_Y$ as that in the first example, which is a smooth function.
We do numerical simulations for  various capillary numbers $|Ca|=0.0025,0.05,0.01$ and $0.02$.
Some numerical results are shown in Figure~\ref{fig:hysteresis}. The first subfigure
corresponds to the case $\delta=0.0003$ and the second is for the case $\delta=0.00003$.
Similar to that in Figure~\ref{fig:hysteresisdelta}, we could see clearly the CAH in
all these cases. We could also see that the advancing contact angle increases 
and the receding contact angle decreases when the absolute value of the capillary number increases(i.e. when the fiber velocity increases). 
More interesting, the dependence of the advancing
and receding contact angles on the velocity are asymmetric in the sense that
the receding angle changes more dramatically than the advancing angle. This can be explained by 
the behaviour of $G(\theta,\cdot)$ as shown in Figure~\ref{fig:steady}.
the slope of the curve  in the receding part is larger than that in the advancing part. 
This implies that 
the receding angle will  change more dramatically than the advancing angle. 
The asymmetric dependence of the CAH on the velocity is consistent with that observed in physical experiments.
The third subfigure in Figure~\ref{fig:hysteresis}, which is taken from \cite{guan2016asymmetric}, shows
the velocity dependence of CAH in experiments.
We can see that the behaviour of the dynamic CAH is very like that in our numerical simulations.
Finally, we would like to remark that similar phenomenon has been observed in a pure phase-field model\cite{XuZhaoWang2019},
without considering the viscous dissipation in fluid. The numerical results by the reduced model~\eqref{e:ODE1} are more consistent with the experiments quantitatively, since viscous dissipations are correctly included by the Onsager principle.

\begin{figure}[ht!]
 \centering
  { 
    \includegraphics[width=2.5in]{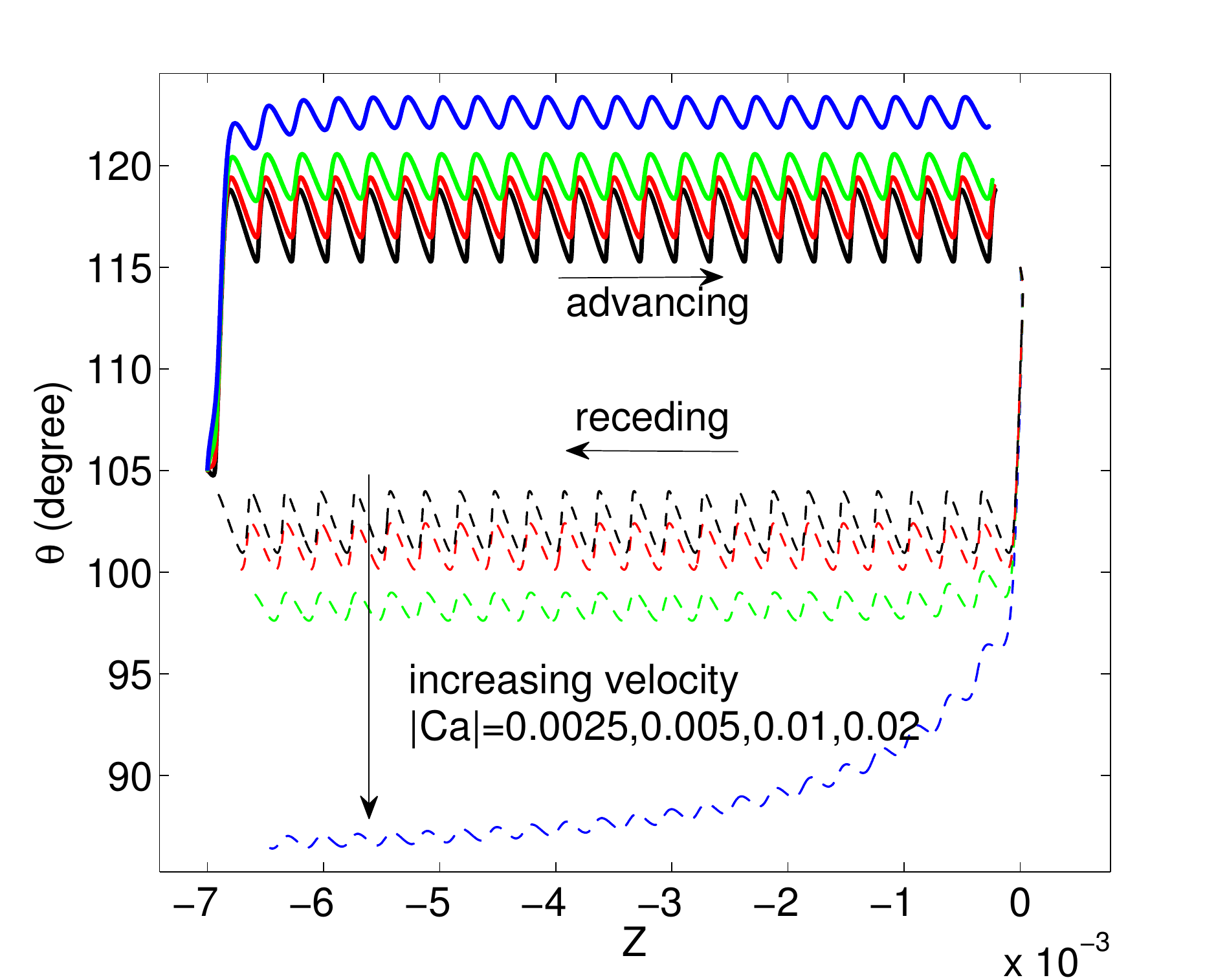}
     \includegraphics[width=2.5in]{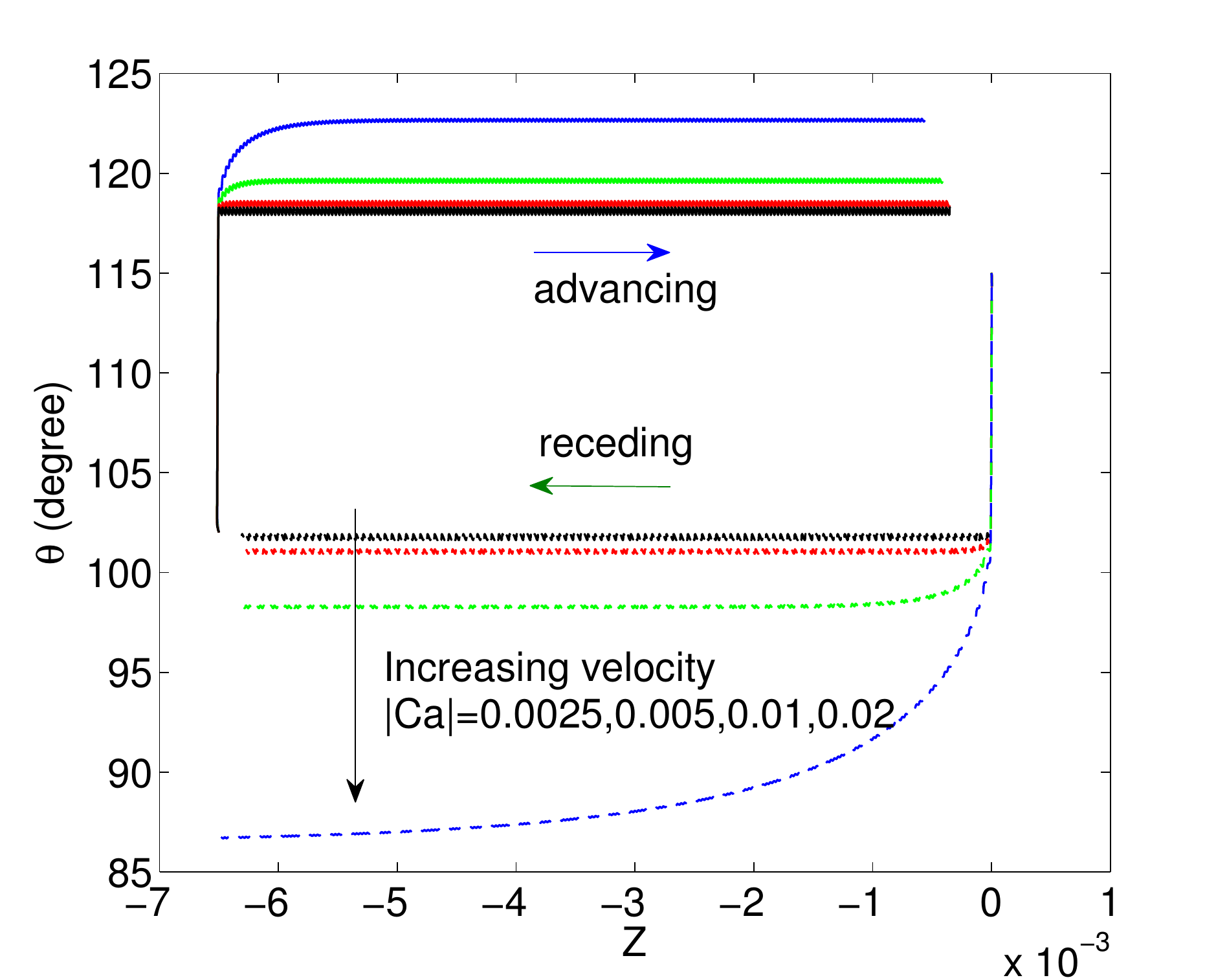}
 \includegraphics[width=3.2in]{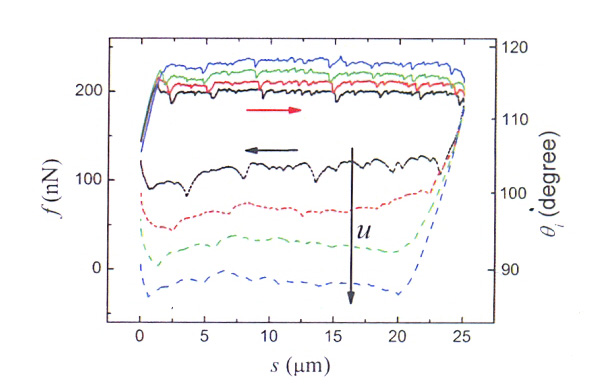}
    }
  \caption{Asymmetry dependence of the advancing and receding contact angles on the wall velocity. The last subfigure shows
 the experimental results in \cite{guan2016asymmetric}}.
  \lbl{fig:hysteresis}
\end{figure}


\section{Conclusions}
We develop a simple model for dynamic CAH on a fiber with chemically inhomogeneous
surface by using the Onsager principle as an approximation tool. 
The model is an ordinary differential system for
the apparent contact angle and the contact line. It  is  easy to analyse and to solve numerically.
The analytical and numerical results show that the model captures the essential phenomena of the CAH. 
By using the model, we derive an upper bound for the 
dynamic advancing contact angle and a lower bound for the receding angle. 
The model is able to characterize  the asymmetric dependence of the advancing 
and receding contact angles on the fiber velocity, which has been observed in 
recent experiments.

Finally, we would like to remark that the real solid surface in experiments is  more complicated than 
the setup in this paper.  The chemical inhomogeneity or geometrical  roughness of the solid surface may be random and the distribution of defects
may lead to very complicated contact line motion. For example, many sophisticated phenomena  have been observed in~\cite{guan2016simultaneous}.
In addition, the heat noise may also affect the relaxation behaviour of the contact line when the radius 
of the fiber is in micro-scale. 
These effects will be studied in future study.


{\bf Acknowledgments:}

The author would like to thank Professor Masao Doi,  Tiezheng Qian, and Penger Tong for their helpful discussions.
This work was supported in part by NSFC grants DMS-11971469
 and  the National Key R\&D Program of China under Grant 2018YFB0704304 and Grant 2018YFB0704300.

{
\bibliography{literW}
}
\end{document}